# Lab Based Curriculum for CIS and Related Technology

Shahriar Movafaghi
Computer Information Technology
*Southern New Hampshire University*
Manchester, USA
s.movafaghi@snhu.edu

Hassan Pournaghshband
College of Computing and Software Engineering
*Kennesaw State University*
Kennesaw, USA
hpournag@kennesaw.edu

## Abstract

The Computer Information System (CIS) is information and communication technology in support of business processes. In this paper, we present a typical undergraduate computer information system curriculum examining the degree of lab intensity and its effect on the course efficacy. A CIS program is usually part of the school of business as it is in support of business processes. We also explore the differences between a CIS curriculum and other computer related technology courses, such as Information Technology (IT), Computer Science (CS), and Software Engineering (SE). The curriculum is composed of several elements such as content and sequence of subjects, classrooms equipped with computer projection, internet, and local network access, and appropriate computing and software infrastructure. We will focus on the importance and adequacy of labs for the CIS curriculum. The proposed CIS curriculum works for a 4-year as well as a 3-year program. This paper provides a recommendation for local and Federal Accreditation agencies and curriculum committees.

## Introduction

An information system is the information and communication technology (ICT) that an organization uses, and includes the way in which people interact with this technology in support of business processes [1]. Information technology (IT) is the application of computers to store, study, retrieve, transmit and manipulate data, [2] or information, often in the context of a business or another enterprise [3]. IT is considered a subset of information and communications technology (ICT). In 2012, Zuppo proposed an ICT hierarchy where each hierarchy level contain[s] some degree of commonality in that they are related to technologies that facilitate the transfer of information and various types of electronically mediated communications [4]. CIS offers more technical courses while IT prepares students for positions such as the position of Chief Information Officer by providing more managerial courses.

Computer Science (SC) is the branch of engineering science that studies (with the aid of computers) computable processes and structures [5]. Computer science programs offer more theoretical courses than information technology. Software Engineering is the application of engineering to the development of software in a systematic method [6][7][8].

The course topics could be the same for all computer technology programs such as computer information systems, information technology, computer science, software engineering, and gaming technology. However, the material covered, and lab requirement and homework assignments usually vary drastically. For example, a database course in computer information systems may include query optimization topics. The student in computer information systems is expected to understand query optimization issues and may be required to explain query optimization concepts, techniques, and issues in a quiz or an exam. In a software engineering course, the students as a group may need to implement a typical query optimizer using SDLC (Software Development Life Cycle). In a computer science course, the students learn theories behind optimization, may be required to read several papers related to query optimization and come up with a new algorithm for the query optimizer. In information technology and gaming courses, the query optimization techniques may be replaced with more managerial topics and gaming databases respectively.

## Typical CIS Curriculum

The information system discipline contributes significantly to several domains, including business and government. Information systems are complex systems requiring both technical and organizational expertise for design, development, and management. They affect not only operations but also the organization's strategy [9]. An effective CIS curriculum is composed of several elements such as the content and sequence of subjects, laboratories, and adequate faculties. The question is how to measure the adequacy of labs and how many courses are needed with appropriate laboratories to support CIS required courses. We have divided the classes with a lab into three categories, namely low, medium and high intensive labs. The low intensive labs are fifty percentage lab related homework and exams, and fifty percent reading, written or multi-choice quizzes. In low intensive lab courses, students usually can install the necessary software on their laptop. The percentage of lab and non-related lab activities in medium intensive lab courses are around 65%-35%, and for a high intensive lab, courses are 75%-25%. It is difficult to install medium intensive lab software on a student's laptop. The physical lab or virtual machine is needed for medium and high intensive lab courses.

Some universities are dismantling CIS labs because of the high cost of supporting the labs and based on the assumption that all CIS students have laptop computers. We agree with this assessment if an appropriate virtual lab is established for courses that require a stable environment. However, we continue to recommend that appropriate physical lab(s) be available for courses that require an unstable environment. In the unstable environment, we may bring down the entire system from time to time for various educational purposes such as demonstrations of web server failover, database failover, clustering, load balancing, reconfiguring partial environment or entire environment, etc. All students have administrative privileges for their own guest operating systems to fulfill their assignments. Some of the courses require that the student have administrative privileges even for the base operating system [10].

Any curriculum should consider recent industry trends. The Gartner group [11] recommendations for CIS related topics is:

- Data scientists understand data and AI algorithms, and formulate coherent questions or problem domains in which to apply these algorithms
- Application developers design interfaces, services and process flows
- The application of advanced big data analytics and AI

The lack of the relevant data sciences will probably hamper AI adoption in the short term. By 2020, 30% of new development projects will deliver AI through joint teams of data scientists and programmers.

Figure 1 shows the required courses of a typical four-year CIS curriculum and related pre-requisites. Figure 2 shows how these courses can be taken in a four-year program.

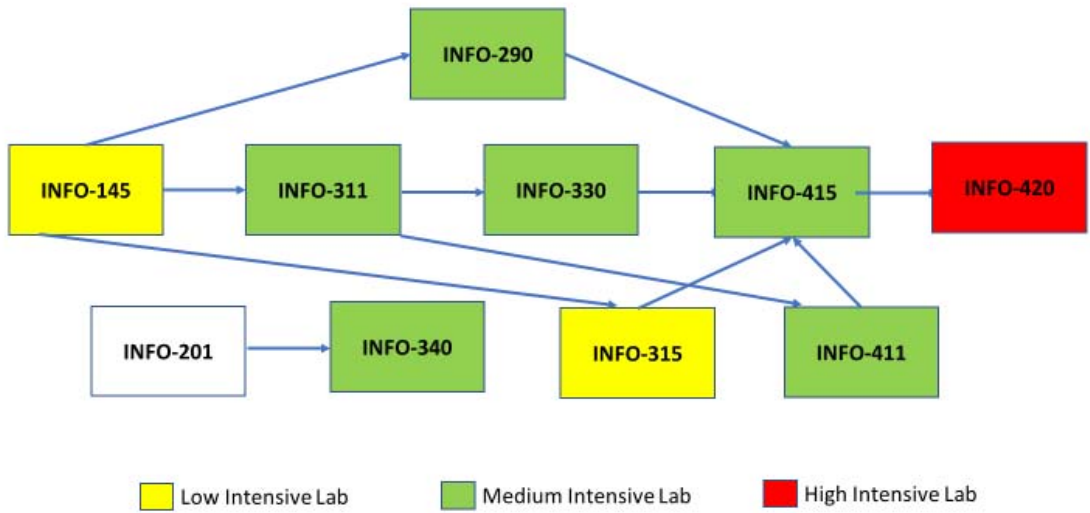

Figure 1 - The required courses of a typical four year CIS curriculum.

| | First Year | | Second Year | | Third Year | | Forth Year | |
|---|---|---|---|---|---|---|---|---|
| | 1st Semester | 2nd Semester | 1st Semester | 2nd Semester | 1st Semester | 2nd Semester | 1st Semester | 2nd Semester |
| | INFO-135 | INFO-145 | INFO-331 | INFO-290 | INFO-330 | INFO-315 | INFO-415 | INFO-420 |
| | | INFO-201 | | | INFO-340 | INFO-411 | | |

Figure 2- Typical CIS required courses in a four year program

The brief description of courses is shown below:

INFO-135 - Interactive Scripting in Virtual Environment
INFO-145 - Introduction to Software Development
INFO-201/INFO-200 - Computer Platform Technologies
INFO-290 – Introduction to Web Services
INFO-311 - Data Structure
INFO-315 - Object Oriented Analysis and Design
INFO-330 - Database Design and Management
INFO-335 - Big Data Fundamentals
INFO-340 - Network and Telecommunication Management
INFO-411 - Artificial Intelligence
INFO-415/INFO-414 - Advanced Information Systems Design (Capstone)
INFO-420/INFO-419 - Advanced Information Systems Implementation (Capstone)

The core courses in the school of business includes many disciplines such courses from economics, finance, and CIS.  Usually, the first course in CIS is a scripting language that is demonstrated as INFO-135.  INFO-145 is the first Java course, and INFO-290 is the second Java course. The course INFO-335 can replace INFO-330 or can be offered as an elective.

Figure 3 shows the same proposed curriculum in three years.  Note that the courses INFO-200, INFO-414, and INFO-419, are the same as INFO-201, INFO-415, and INFO-420 except that there is no class.

| First Year | | Second Year | | Third Year | |
|---|---|---|---|---|---|
| 1st Semester | 2nd Semester | 1st Semester | 2nd Semester | 1st Semester | 2nd Semester |
| INFO-135 | INFO-145 | INFO-290 | INFO-330 | INFO-414 | INFO-419 |
| INFO-200 | INFO-339 | INFO-311<br>INFO-315 | INFO-411 | | |

Figure 3- Typical CIS required courses in a three year program

## Recommendation

Curriculum creation and modification is the responsibility of the relevant faculty. The curriculum approved by the faculty is then passed to school dean, school curriculum committee, university curriculum committee, and provost for approval. Unfortunately, many faculty and administrators are not aware of the fact that labs are part of the curriculum requirement. Further, some faculty produce an entire curriculum or part of a curriculum using templates, which makes the requirement for the lab difficult, if not impossible. Below is a list of recommendations for faculty, university administrators, local and federal accreditation agencies:

1. All faculty, university administrators, local and federal accreditation agencies comments should be logged and be accessible to all the parties involved. If an automated tool such as Kuali Curriculum Management System [2] is used, then the faculty vote, and comments should be recorded.

2. The courses delivered to the curriculum committee should include the lab requirements.

3. A VM (virtual machine) is recommended for all the students in a CIS program. It would be desirable to give a VM to all of the students at the beginning of the four-year or three-year program that includes all necessary software for the CIS courses. this way, students can store their data on a separate disk so that the CIS department can update the software in the VM without affecting the student's data.

4. The university, local and federal accreditation agencies should make a more rigorous assessment of the quality of labs and quality of lab classes in CIS and related technologies.

5. Adequately trained and experienced faculty is also a part of the curriculum. The evaluation of faculty members by other faculty members should also be recorded, even if it is anonymous, for local and federal accreditation agencies.

## Conclusion and Future Research

In this paper, we proposed a four-year and three-year lab-based typical CIS curriculum. We established a criterion to evaluate CIS labs. We divided the classes with the lab into three categories, namely, low, medium and high intensive lab classes. The proposed CIS curriculum is based on the trend of technology. The list of recommendation for faculty, university administrators, local and federal accreditation agencies is provided. Future research includes examining the local and federal accreditation process so that we can improve their assessments regarding the quality of the curriculum.